\documentclass[prd,preprint,eqsecnum,nofootinbib,amsmath,amssymb,
               tightenlines,dvips]{revtex4}
\input epsf
\usepackage{epsfig}
\usepackage{bm}

\def\centerbox#1#2{\centerline{\epsfxsize=#1\textwidth\epsfbox{#2}}}
\def\Tr{\,{\rm Tr}\:}
\def\commut#1#2{\Big[#1\,,\, #2\Big]}
\def\st{\begin{equation}}
\def\stp{\end{equation}}
\def\sss{\scriptscriptstyle}
\def\brem{bremsstrahlung}
\def\lra{\leftrightarrow}

\def\Re{\,{\rm Re}\:}
\def\Im{\,{\rm Im}\:}

\def\Eq#1{Eq.~(\ref{#1})}

\def\twotwo{{2\lra2}}

\def\j{{\bm j}}
\def\x{{\bm x}}
\def\p{{\bm p}}
\def\q{{\bm q}}
\def\k{{\bm k}}
\def\v{{\bm v}}

\def\E{{\bm E}}

\def\F{{\bm F}}

\def\cf{C_{\rm F}}

\def\nf{N_{\rm f}\,}
\def\nc{N_{\rm c}\,}

\def\mD{m_{\rm D}}

\def\trans{\top}

\newcommand\ansatz{{\it Ansatz}}

\def\half{{\textstyle{\frac 12}}}

\def\alphas{\alpha_{\rm s}}

\def\nott#1{\setbox0=\hbox{$#1$}                
   \dimen0=\wd0                                 
   \setbox1=\hbox{/} \dimen1=\wd1               
   \ifdim\dimen0>\dimen1                        
      \rlap{\hbox to \dimen0{\hfil/\hfil}}      
      #1                                        
   \else                                        
      \rlap{\hbox to \dimen1{\hfil$#1$\hfil}}   
      /                                         
   \fi}                                         %

\def\lsim{\mbox{~{\raisebox{0.4ex}{$<$}}\hspace{-1.1em}
	{\raisebox{-0.6ex}{$\sim$}}~}}

\advance\parskip 2pt

\begin {document}


\title
    {
      Dileptons, spectral weights, and conductivity in the Quark-Gluon
      Plasma
    }

\author{Guy D.~Moore and Jean-Marie Robert}
\affiliation
    {%
    Department of Physics,
    McGill University, 
    3600 University St.,
    Montr\'{e}al, QC H3A 2T8, Canada
    }%
\date {February 2003}

\begin {abstract}%
    {%
    We re-examine soft dilepton emission from a weakly coupled
    Quark-Gluon Plasma.
    We show that Braaten, Pisarski, and Yuan's result that the dilepton
    rate rises as $1/q_0^4$ (and the spectral weight scales as $1/q^0$) at
    small energy $q^0 \ll gT$ is correct, but
    that the coefficient they found for this behavior is not correct,
    because their analysis was incomplete.  At still smaller scales, the
    behavior moderates to $\propto 1/q_0^2$ for $q^0 \lsim g^4 T$,
    consistent with a finite electrical conductivity.  We evaluate the
    spectral weight in the $q^0 \sim g^4 T$ region by kinetic theory
    techniques and show that it satisfies a sum rule, which makes the
    determination of electrical conductivity from the Euclidean
    correlation function very challenging.
    }%
\end {abstract}

\maketitle

\section {Introduction}

It is believed that experiments at RHIC are producing the Quark-Gluon
Plasma, which will be further probed by the heavy ion program at the
LHC.  However, one of the complications in extracting the properties of
the Quark-Gluon Plasma is that thermalization and rescattering tends to
erase information about the state of the plasma at early times.
Therefore, there has been an effort to understand so-called ``hard
probes,'' particles which are produced early during a heavy ion
collision and either escape without further interaction, or interact 
with the high density plasma much more strongly than with the hadrons
present at late times.  Jets and heavy quarks are examples of the
latter; photons and dileptons are examples of the former.  For a
comprehensive review, see for instance \cite{yellowbook}.

The thermal production of dileptons has long been viewed as a promising
means to identify and study the Quark-Gluon Plasma
\cite{dilepton_early,Toimela}.  Theoretically, our two best tools for
establishing the dilepton production rate by the plasma are perturbation
theory and lattice gauge theory.  Each is rigorous within some regime.
Both suffer from limitations.  In particular, while perturbative tools
are rigorous in the limit of small coupling (for a fixed energy), they
probably do not show good convergence at the value of coupling constants
relevant in realistic experiments.  Furthermore, the application of
perturbation theory has proven much more complicated than naively
expected.  In particular, when momenta are small (or close to the light
cone \cite{Rebhan}), the perturbative expansion requires a resummation,
the Hard Thermal Loop (HTL) resummation, pioneered by Braaten and Pisarski
\cite{BraatenPisarski,Taylor}.

The thermal dilepton production rate appears to be under control for
large, deeply timelike momenta.  Perturbation theory works well; the
leading and $O(\alphas)$ subleading behavior is known, and there are no
singularities encountered \cite{Majumder}.  Further, lattice gauge
theory techniques are
applicable and show good agreement with perturbation theory
\cite{Petreczky}.  However, the situation is less clear for small or
nearly lightlike momenta.  For large but nearly lightlike
momenta, perturbation theory can be applied, but it requires rather
extensive resummation \cite{AGMZ}.  It may be quite difficult to compare
lattice techniques to the perturbative results in this regime, because
the lattice only determines
correlation functions at Euclidean time separations; an analytic
continuation by the Maximal Entropy Method \cite{MEM} is required to establish
dilepton production rates.  This method does not work well where the
relevant correlation function varies rapidly with frequency, which
occurs near the light cone.

The problem at low frequencies is also confused.  In one of the
pioneering papers of the HTL resummation technique, Braaten, Pisarski,
and Yuan (BPY) computed the small energy, zero momentum
dilepton production rate, using HTL resummation to obtain the full
leading behavior \cite{BPY}.  However, lattice results \cite{Petreczky,Gupta}
do not seem to agree with their results.  Arguably, neither the
perturbative result, nor the results from the Bielefeld lattice group
\cite{Petreczky}, can
be correct, at least in the very low energy limit.  The low energy
limit of the dilepton rate is simply related to the electrical
conductivity of the Quark-Gluon Plasma, as we will review below.  On
physical grounds, the electrical conductivity should be nonzero and
finite.  However, the lattice results from the Bielefeld group seem to
extrapolate to a
conductivity which is zero--while the results of BPY
extrapolate to a conductivity which is infinite.

This paper will re-analyze the low energy region, using weak coupling
techniques.  BPY found a dilepton production
rate $dW/d^4 q \sim q_0^{-4}$, corresponding to a current-current
correlator spectral
weight $\rho(Q) \sim 1/q^0$, for $q^0$ smaller than but formally of order
$gT$.  We give a clear physical picture of how this scaling behavior
arises, and then show that it breaks down at the scale
$q^0 \sim g^4 T$, formally outside the domain of validity of the Braaten,
Pisarski and Yuan calculation.  At very small $q^0$, $\rho/q^0$ takes a
shape similar to, though distinct from, a Lorentzian, so
$\lim_{q^0 \rightarrow 0} \rho/q^0$ is finite, consistent with a finite
electrical conductivity.  Our quantitative results in this region can be
considered an extension of existing calculations of the electrical
conductivity \cite{AMY1,AMY6}.

More surprisingly, we find that even in the claimed region of validity
of Braaten, Pisarski, and Yuan's calculation, $q^0 \sim gT$, their result
for the dilepton production rate
is incomplete; for $q^0 \ll gT$, it is low by about a factor of about 4.
This is because the power counting underlying their calculation is flawed;
there are diagrams which were discarded in their analysis which in fact
contribute at the same order in the coupling as those they have
included.  We provide a complete calculation for the case $q^0 \ll gT$.

\section{Spectral weight, conductivity, dilepton rate}

First let us review the relation between the current-current spectral
weight, the conductivity, and the dilepton production rate.  
Dilepton pairs are produced electromagnetically.  The amplitude
arises at second order in the electromagnetic interaction,
requiring one insertion of $eA_\mu J^\mu_{\rm lept}$, with
$J^\mu_{\rm lept}$ the leptonic electromagnetic current, and one
insertion of $A_\nu J^\nu_{\sss QCD}$ with
$J^\nu_{\sss QCD}$ the hadronic electromagnetic current,
\st
J_{\sss QCD}^\mu = \sum_q Q_q i \bar{q} \gamma^\mu q \, .
\stp
(We use $[{-}{+}{+}{+}]$ metric convention.)
The differential rate to produce a dilepton pair of 4-momentum $q^\mu$
from a QCD plasma described by a density matrix $\rho$
depends on the squared matrix element,
\begin{eqnarray}
\sum_{h}\int_{prr'} \frac{1}{Z}\Tr \rho \;
eJ^\alpha_{\sss QCD}(q) A_\mu(-q) \; eJ^\beta_{\rm lept}(-r') A_\nu(r') 
\; |\, l^+(p)l^-(q-p) h \rangle\nonumber
\\
\times
\langle l^+(p) l^-(q-p) h\, |\;
eJ^\mu_{\sss QCD}(-q) A_\mu(q) \;eJ^\nu_{\rm lept}(r) A_\nu(-r) 
\, ,
\end{eqnarray}
where $|h\rangle$ represents an arbitrary hadronic state.
The gauge fields contract to give propagators, 
$\eta_{\mu\nu}\delta(q-r)/q^2$, and the leptonic part of the
calculation can be evaluated perturbatively \cite{Toimela}.
The sum on the hadronic final state $|h\rangle$ is the identity.
Performing the leptonic part of the calculation, the
differential dilepton production rate per unit 4-volume is
\st
\frac{dW(q)}{d^4 q} = \frac{\alpha^2}{6\pi^3 q^2}
\int_{q'} \frac{1}{Z} \Tr \rho J_\mu(q') J^\mu(-q)
= \frac{\alpha^2}{6\pi^3 q^2}
\int d^4 x e^{iq\cdot x} \frac{1}{Z} \Tr \rho J_\mu(x) J^\mu(0) \, ,
\stp
where $W(q) d^4 q$ represents the production rate per 4-volume of
dileptons of 4-momentum $q$.
The required correlation function is called the Wightman
current-current correlation function,
\st
\tilde\Pi^{<}_{\mu\nu}(q) \equiv 
\int d^4 x e^{iq\cdot x} \frac{1}{Z} \Tr \rho J_\mu(x) J_\nu(0)
 = \int d^4 x e^{iq\cdot x}
\langle J_\mu (x)J_\nu (0)\rangle_\beta \,.
\stp 
The current correlation is proportional, at lowest order in e and to all
orders in g, to the one-particle irreducible photon self-energy,
$e^2\tilde\Pi^{<}_{\mu\nu}(q)= \Pi^{<}_{\mu\nu}(q)$. 

The spectral weight is defined using the commutator of the current
operators, 
\st
\rho_{\mu\nu}(q) \equiv \int d^4 x e^{iq\cdot x}
\left\langle \commut{J_\mu(0)}{J_\nu(x)} \right\rangle_\beta
= -\tilde\Pi^{<}_{\mu\nu}(q) + \tilde\Pi^{<}_{\nu\mu}(-q) \,.
\stp
In a general density matrix there need not be a simple relation between
$\rho_{\mu\nu}(q)$ and $\tilde\Pi^{<}(q)$, but for a thermal density
matrix, they are related by the KMS condition \cite{KMS}:
\st
\rho_{\mu\nu}(q) = \left(e^{\beta q^0} - 1\right) 
\tilde\Pi^{<}_{\mu\nu}(q) \,,
\quad \mbox{or} \quad
\tilde\Pi^{<}_{\mu\nu}(q) = \frac{1}{e^{\beta q^0}-1} \rho_{\mu\nu}(q)
=n_b(q^0) \rho_{\mu\nu}(q) \, ,
\label{eq:rho}
\stp
with $n_b$ the Bose statistical function.  Note, for $q^0<0$, this
relation is, $\rho_{\mu\nu}(q)=-[1{+}n_b(|q^0|)]\tilde\Pi^{<}_{\mu\nu}(q)$.
Therefore, in terms of the spectral weight, the dilepton production rate
is
\st \label{DileptonRate}
\frac{dW(q)}{d^4 q} = \frac{\alpha^2
  \rho(q)}{6\pi^3q^2(e^{\beta q_0} -1)} \, ,
\stp
where $\rho(q) = g^{\mu\nu}\rho_{\mu\nu}(q)$.

Note that the spectral weight $\rho_{\mu\nu}(q)$ is odd in
$q$, $\rho_{\mu\nu}(q)=-\rho_{\mu\nu}(-q)$.  Also, its
relation to the (well behaved) Wightman correlator,
\Eq{eq:rho}, means that $\rho(q)$ will vanish linearly as
$q^0\rightarrow 0$.  This sometimes makes working in terms of $\rho$
awkward; it is $\rho_{\mu\nu}(q)/q^0$ which is well behaved near
$q^0=0$.  In particular, consider the zero momentum limit of the
time-time spectral weight, $\rho_{00}(\vec{q}=0,q^0)$.
It is related to the Wightman correlator
$\tilde\Pi^{<}_{00}(\vec{q}=0,q^0)$, which is the frequency transform of
the real-time Wightman correlation function,
\st
\tilde\Pi^{<}(\vec{q}=0,t) = 
\int d^3 x \tilde\Pi^{<}_{00}(x,t) = \int d^3 x \langle J_0(x,t)
J_0(0,0) \rangle \, .
\stp
But this is just the charge susceptibility; writing the spatial volume
as $V$, and using translation invariance,
\st
\int d^3 x \tilde\Pi^{<}_{00}(x,t) = \int d^3 x \langle J_0(x,t)
J_0(0,0) \rangle = \frac{1}{V} \int d^3 x d^3 y \langle J_0(x,t)
J_0(y,0) \rangle = \frac{1}{V} \langle Q(t) Q(0) \rangle \, ,
\stp
with $Q=\int J_0 d^3 x$ the total charge.  Since charge is conserved,
$Q(t)=Q(0)$.  The susceptibility is defined as,
\st
\chi_Q \equiv \frac{1}{V} \langle Q^2 \rangle \, ,
\stp
so clearly
\st
\tilde\Pi^{<}_{00}(\vec{q}=0,t) = \chi_Q \, .
\stp
Fourier transforming,
\st
\tilde\Pi^{<}_{00}(\vec{q}=0,q^0) = \chi_Q 2\pi \delta(q^0) \, ;
\stp
the correlator vanishes away from $q^0=0$ but has a delta function
there.  The spectral weight $\rho_{00}$ will also vanish away from zero,
but the spectral weight is multiplied by $1/n_b(q^0) = 0$.  However,
$\rho/q^0$ retains the delta function and ensures that frequency
integrals involving $\rho_{00}$ will correctly be proportional to the
charge susceptibility.

Now, consider electrical conductivity.  It is defined in terms of the
current response induced by a static electrical field,
\st
\langle e J_i^{EM} \rangle = \sigma \langle E \rangle,
\stp
where $\langle E \rangle$ is a small, externally imposed electrical
field.  We assume here that we are in the local fluid
rest frame. 


The value of the conductivity can be related to the current-current
correlator discussed earlier via a Kubo relation \cite{Kubo}.
Essentially, an external electric field, 
$E_i(t) = E_i \Re e^{-iq^0 t}$, is added to the Hamiltonian.  This
involves introducing a $J_i A_i$ term, with $E_i(t)=\partial_0 A_i$, as
a perturbation to the Hamiltonian;
\st
H(t) = H_0 + H_I\, , \quad
H_I = \int d^3 x \; e A_i(x,t) J_i(x,t) \, , \qquad
A_i = \frac{E_i}{-iq^0} \, ,
\stp
where the real part is to be taken at the end.  Here $H_0$ is the {\em
  full} QCD Hamiltonian, but neglecting electromagnetic interactions and
without any external electrical field.
The current is evaluated in this perturbed system, fixing retarded
boundary conditions such that the initial conditions at $t=-\infty$ are
the equilibrium density matrix.  The conductivity is given by the ratio
of current to electric field in the small frequency limit.  Since we are
interested in weak electrical fields, we may work in the interaction
picture and expand to first order in the perturbation;
\begin{eqnarray}
\langle J_i(t=0) \rangle_{\rm pert} & = & \Re \Tr \rho_{\rm th} \;
\left( \exp \int_{-\infty}^{0} -i H_I(t') dt' \right)^\dagger J_i
\left( \exp \int_{-\infty}^{0} -i H_I(t') dt' \right)
\nonumber \\
& = & \Re \int_{-\infty}^{0} dt' \Tr \rho_{\rm th} \left( iH_I(t') J_i
-J_i H_I(t') \right)
\nonumber \\
& = & \Re \int d^3 x \int dt' \Theta(-t') \Tr \rho_{\rm th}
-ie \commut{J_i(0,0)}{A_j(x,t') J_j(x,t')} \,.
\end{eqnarray}
Since $A_j$ is an external field, we can pull it out, leaving the
current-current commutator.  Also substituting $A$ for $E$ and dividing
it through, we find \cite{Hosoya},
\st
\frac{\langle eJ_i \rangle}{E_j} = \frac{e^2}{q^0} 
\Re \tilde\Pi^{\rm R}_{ij}(\vec q=0,q^0) \, ,
\label{PI_ret}
\stp
where $\tilde\Pi^{\rm R}$ is the {\em retarded} correlator, the one
including the $\Theta(t)$ time ordering step function.  Its real part is
related to the spectral weight by%
\footnote{%
    The spectral weight is the discontinuity of the propagator across
    the real axis, equivalent to the sum of the retarded and
    advanced propagators.  When time translation invariance holds,
    as in equilibrium, the retarded and advanced propagators are related
    as $G_{\rm A}=-G^*_{\rm R}$, so $\rho=2\Re G_{\rm R}$.
    The retarded correlator is frequently introduced with a factor of $i$
    in the definition, in which case it would be twice the imaginary part
    which equals the spectral weight.
    }%
\st \label{RhoReal}
\rho_{\mu\nu}(q) = 2 \Re \tilde\Pi^{\rm R}_{\mu\nu}(q) \, .
\stp
Rotational
invariance of the plasma ensures that $\rho_{ij}(\vec q=0)
= \frac{1}{3} \delta_{ij} \rho_{kk}(\vec q=0)$, and so
\st
\sigma = \frac{e^2}{6} \lim_{q^0 \to 0} \frac{1}{q^0}
\rho_{ii}(\vec q=0,q^0) \, .
\stp
This is related to the Wightman correlator introduced previously by
\st
\sigma = \frac{e^2}{6} \lim_{q^0 \rightarrow 0}
\frac{\rho_{ii}(\vec{q}=0,q^0)}{q^0}
= \frac{\beta e^2}{6} \lim_{q^0 \rightarrow 0} 
\tilde\Pi^{<}_{ii}(\vec{q}=0,q^0)\, .
\stp

It is expected that in a real, dissipative system such as QCD at finite
coupling, the electrical conductivity should be nonzero and finite.
Therefore the small $q^0$ limit of $\rho_{ii}$ should be $\rho_{ii}
\propto q^0$, and the limit of $W(q)$ should be $W(q)\propto
(q^0)^{-2}$.

The weak coupling expansion for such correlation functions near zero
frequency turns out to
be quite involved \cite{Jeon,Basagoiti,Aarts}; nevertheless,
leading-order weak-coupling results for the conductivity of QCD exist
\cite{AMY1,AMY6}.

\section{Spectral Weight and Kinetic Theory}

Our emphasis in this paper is on the spectral weight at zero momentum
and at frequencies $q^0 < gT$.  The scale $gT$ is relevant because
the dominant scattering process in the plasma, soft gluon exchange,
involves energy and momentum exchanges of order $gT$ (since lower
exchange momenta are screened).  Therefore, the
physical time scale for such a scattering is $\sim 1/gT$, parametrically
shorter than the $1/q^0$ time scale separating the current operators
in the spectral weight.  Therefore, to the extent that one is willing to
perform an expansion in $q^0/gT \ll 1$, one can take the scattering
events to be instantaneous.  In this case, the current response can be
calculated using kinetic theory.

The way to use kinetic theory to find the spectral weight is to take the
derivation of the Kubo relation, discussed above, and use it in reverse;
one computes the spectral weight by evaluating the current in a theory
with an externally imposed electrical field.  Because the weakly coupled
theory has well defined quasiparticles, and because they dominate the
current, energy density, and so forth, a kinetic theory (Boltzmann
equation) description should capture the leading order behavior of the
current in the electric field background.

We first introduce a distribution function, $f(\p ,\x , t)$,
characterizing the phase space density of the quasi-particles.%
\footnote{%
    Our convention is that $f$ is the phase space density for a given
    spin and color, not summed over spins and colors.
    }
This function satisfies the Boltzmann equation,
\st \label{Boltzmann}
\Bigg[\frac{\partial}{\partial t} + \v_\p \cdot \frac{\partial}{\partial
    \x} + \F_{\rm ext} \cdot\frac{\partial}{\partial \p}\Bigg] f(\p ,\x , t) =
-C[f] \, ,
\stp
where $\v_\p \equiv \partial p^0 / \partial \p$ is the velocity of an
excitation with momentum $\p$, $\F_{\rm ext}$ is an external force acting
on the excitation, and $C[f]$ is the collision integral which
characterizes scatterings with other excitations in the plasma.  Since we
are primarily concerned with weakly coupled QCD at small or vanishing
quark mass, we can take the velocities to be lightlike,
$\v_\p = \hat\p$, with $m^2/T^2$ and $g^2$ corrections which we will
neglect.  

We are interested in the case where everything is
statistically spatially homogeneous, but $\F_{\rm ext} = Q e \E$ is
determined by an externally imposed electric field.
For this case, the lefthand side of the Boltzmann equation can be
evaluated, if we linearize in the electric field (which is all that is
necessary to obtain the current-current correlation function).  At
linearized order, we can write
\st
f(\p,\x,t) = f_0(\p) + \delta\! f(\p,t) \, , \qquad
\delta\! f(\p,t) = \delta\! f(\p) e^{-iq^0 t} \, ,
\stp
with $\delta\! f(\p)$ in general complex.  (When we take the real part
of the retarded correlator to get the spectral weight, only the real
part of $\delta\! f$ will contribute.)
Since the electric field is already linear order in $\E$, the term
containing $\E$ need only be expanded to zero order in the electric
field,
\begin{eqnarray}
\F_{\rm ext} \cdot \frac{\partial}{\partial \p} f_0(\p)
& = & Qe \E \cdot \frac{\partial}{\partial \p} 
    \left[ e^{|\p|/T}\mp 1\right]^{-1}
\nonumber \\
& = & \frac{-Qe \E \cdot \hat{\p}}{T} f_0(1{\pm} f_0) \, , 
\end{eqnarray}
where the upper sign is for bosons and the lower sign is for fermions.
In fact, since the electric charge $Q=0$ for gluons, only the lower sign
is needed.

The time derivative acts only on $\delta\! f$, giving
\st
\partial_t f(\p,t) = -iq^0 \delta\! f(\p,t) \, .
\stp
Since we know that $\delta\! f$ arises at linear order in the electric
field, it must be proportional to $\E$.  Rotational symmetry completely
fixes the possible form of this angular dependence, 
$\delta\! f(\p) \propto \E \cdot \hat{\p} \;\delta\! f(p)$, with
$\delta\! f(p)$ an unknown function of $p=|\p|$ only.  It is convenient
to parametrize this unknown function as
\st
\delta\! f(\p) = f_0(1{-}f_0) f_1(\p) = 
\frac{Qe\E \cdot \hat{\p}}{T^2} f_0(1{-}f_0) \chi(p) \, ,
\stp
with $\chi(p)$ unknown and to be determined.

The collision term represents the change in the number of particles of
momentum $\p$ due to scatterings and is a sum of two terms; a loss term
represents the scattering of a particle with momentum $\p$, out of this
momentum state; the gain term accounts for scatterings which generate a
particle of momentum $\p$.  For $2\leftrightarrow 2$ scattering
processes, the collision term is
\begin{eqnarray}\label{C22}
C_a^{2\leftrightarrow 2}[f](\p) &=&
\frac{1}{4|\p|\nu_a}\sum_{bcd}\int_{\k\p'\k'}
|\mathcal{M}^{ab}_{cd}(\p,\k;\p',\k')|^2(2\pi)^4\delta^{(4)}(P+K-P'-K')
 \\
&\times& \Big\{ f^a(\p)f^b(\k)[1\pm f^c(\p')][1\pm
      f^d(\k')]
- f^c(\p')f^d(\k')[1\pm f^a(\p)][1\pm
      f^b(\k)]\Big\} \, , \nonumber
\end{eqnarray}
where the first product of statistical functions is the loss term and
the second is the gain term.%
\footnote{%
    We write the collision term as loss minus gain, and put a $-$ sign
    in front of it in \Eq{Boltzmann}, so that its linearization will be
    a positive operator.
    }
Here $|\mathcal{M}^{ab}_{cd}|^2$ is the spin and color summed squared
matrix element for $ab \rightarrow cd$ and 
$\int_\k \equiv \int d^3\k/(2\pi)^3 2k^0$ is the Lorentz invariant
phase space integral.  The leading factor $1/4p^0 \nu_a$ is the usual
$1/2p^0$, a $1/\nu_a$ to replace the initial spin and color summed
$|\mathcal{M}|^2$ with the spin and color averaged one, and a $1/2$
to eliminate double counting final states, $cd\rightarrow dc$ (or as
a final state symmetry factor when $c=d$).

Just as we have linearized the lefthand side of the Boltzmann equation,
now we must linearize the collision term.  Because of our particularly
convenient choice $\delta\! f = f_0(1\pm f_0) f_1$, the loss minus gain
difference of population functions takes a very simple form,
\begin{eqnarray}
&&
\Big\{ f^a(\p)f^b(\k)[1{\pm} f^c(\p')][1{\pm}f^d(\k')]
- f^c(\p')f^d(\k')[1{\pm} f^a(\p)][1{\pm}f^b(\k)]
\Big\}
\nonumber \\
& = & f^a_0(\p) f^b_0(\k) [1\pm f^c_0(\p')][1\pm f^d_0(\k')]
\left\{ f_1(\p){+}f_1(\k){-}f_1(\p'){-}f_1(\k') \right\} \, .
\end{eqnarray}
Substituting
\st
f_1(\k) = Qe\frac{\E\cdot \hat\k}{T^2} \chi(k)
\stp
looks like it will make the result complicated; on the left side of the
Boltzmann equation, only $\E\cdot \hat{\p}$ appears.  Fortunately, the
phase space, matrix element, and equilibrium distribution functions are
all invariant on rotations about the $\hat\p$ axis.  Averaging over such
rotations replaces $\hat\k$ with $(\hat\k\cdot \hat\p) \hat\p$, so after
such averaging, the Boltzmann equation becomes
\begin{eqnarray}
QT f_0(1-f_0)\hat \p + iq^0 Q f_0(1-f_0)\hat \p \chi(p) 
= \frac{\hat\p}{4|\p|\nu_a}\sum_{bcd}\int_{\k\p'\k'}
|\mathcal{M}^{ab}_{cd}(\p,\k;\p',\k')|^2
\nonumber\\
\times(2\pi)^4\delta^{(4)}(P+K-P'-K')
f^a_0(\p) f^b_0(\k) [1\pm f^c_0(\p')][1\pm f^d_0(\k')]
\nonumber\\
\times
\hat\p \cdot 
\left[Q\hat \p \chi(p)+Q_b\hat \k\chi(k)
  -Q_c\hat \p'\chi(p')-Q_d\hat \k'\chi(k')\right]\, ,
\label{eq:Boltzmann_first}
\end{eqnarray}
or
\begin{eqnarray}
QT & = & -iq^0 Q \chi(p) + \frac{1}{f_0(1{-}f_0)}
\frac{1}{4p\nu_a} \sum_{bcd} \int_{\k\p'\k'}
|\mathcal{M}^{ab}_{cd}(\p,\k;\p',\k')|^2
(2\pi)^4\delta^{(4)}(P{+}K{-}P'{-}K')
\nonumber \\ && \hspace{1.8in} \times
f^a_0(\p) f^b_0(\k) [1\pm f^c_0(\p')][1\pm f^d_0(\k')]
 \\ && \hspace{1.8in} \times
\left[Q\chi(p)+\hat\p \cdot \hat \k Q_b \chi(k)
  -\hat\p \cdot \hat\p' Q_c \chi(p')
  -\hat\p \cdot \hat\k' Q_d \chi(k')\right]\, . \nonumber
\label{eq:Boltzmann_full}
\end{eqnarray}
Symbolically, this can be expressed as,
\begin{equation}
S(\p) = \left[ -iq^0 + \hat{\cal C} \right] \chi(\p) \, ,
\label{eq:Boltzmann_symbols}
\end{equation}
where $\hat {\cal C}$, the linearized collision operator defined above,
is a nonlocal operator in $\p$ space.  This can be formally solved for
$\chi$ by inversion, $\chi(\p) = [-iq^0 + \hat{\cal C}]^{-1} S(\p)$.

We will solve this (integral) equation in subsequent sections.  First,
though, we show what the solution has to do with the current-current
correlation function which we are after.  The current, in the kinetic
theory approximation, is
\begin{eqnarray}
j_i & = & \sum_a \nu_a \int_\p Q_a e \hat\p_i f_a(\p,x)
\nonumber \\
 & = & \sum_a \nu_a \int_\p Q_a e \hat\p_i f_0(1{-}f_0) \,\delta\! f_a(\p,x)
\nonumber \\
 & = & \sum_a \nu_a \int_\p Q_a^2 e^2 \hat\p_i \frac{\E\cdot \hat\p}{T^2}
	f_0(1{-}f_0) \chi_a(p)
\nonumber \\
 & = & \sum_a \nu_a Q_a^2 e^2 \frac{\E_i}{3T^2} \int_\p f_0(1{-}f_0) 
        \chi_a(p) \, .
\end{eqnarray}
In passing to the final line we used that the angular average of
$\hat\p_i \hat\p_j$ is $\delta_{ij}/3$.  We see as expected that $\j
\propto \E$; (the integral of) the function $\chi$ expresses the
coefficient of this proportionality.  Since $\chi$ is in general complex
(see \Eq{eq:Boltzmann_symbols}), the current typically lags the electric
field; the dot product $\j\cdot \E/\E^2$ depends on the real part of
$\chi(p)$.  Using \Eq{PI_ret} and \Eq{RhoReal} from the last section, we
see that the spectral weight is
\begin{equation}
\frac{\rho_{ij}(q^0)}{q^0} = \frac{2\delta_{ij}}{3T}
        \sum_a \nu_a Q_a^2 \int_\p f_0(1{-}f_0) \chi_a(p) \, ,
\qquad
\frac{\rho}{q^0}=2\sum_a \nu_a \frac{Q_a^2}{T} 
	\int_\p f_0(1{-}f_0)\chi_a(p)\,.
\label{eq:rho_j}
\end{equation}
Therefore the finite frequency solution to the Boltzmann equation yields
the spectral weight at finite frequency.
This is a generalization of the technique used to
determine the electrical conductivity in \cite{AMY1}.

\section{Diagrams, photon emission, and Boltzmann equation}

Before computing the spectral weight, we pause to discuss the relation
between what we have just done and the usual diagrammatic approach to
the photon production rate.  In doing this, we will see the physical
origin of the flaw in the treatment of BPY.

We begin by emphasizing the range of validity of each calculation.  The
Boltzmann calculation is most naturally phrased in terms of frequencies
$q^0 \sim g^4 T$, since this is where the collision and $iq^0$
terms in the Boltzmann equation, \Eq{eq:Boltzmann_symbols}, are of the
same order.  However, the treatment only breaks down (at weak coupling)
when the ``instantaneous scattering'' approximation in the Boltzmann
treatment ceases to be valid, so it is valid provided $q^0 \ll gT$.

The treatment of BPY was billed as a complete
calculation for $q^0 \sim gT$.  As such, it should be valid over a
parametrically wide range.  By treating $q^0$ as formally $O(gT)$,
the authors' approximations could become uncontrolled when $q^0$
differs from this estimate by a power of $g$, for instance, for $q^0
\sim T$ or $q^0 \sim g^2 T$.  However, the region 
$g^2 T \ll q^0 \ll gT$ should be within the validity of their
approximations.  The two calculations overlap in this range, and the
absence of any change in behavior at the $g^2 T$ scale in the Boltzmann
treatment shows that the BPY calculation should in
fact be valid for any scale $q^0 \gg g^4 T$.  However, their
treatment certainly fails below this scale (and they never claimed
otherwise).  Hence, their treatment is valid for
$g^4 T \ll q^0 \ll T$.
Therefore, both treatments should be valid in the range
$g^4 T \ll q^0 \ll gT$, and we can compare with their
results in this regime.

Because $q^0 \gg {\cal C}$ in this regime, we
can simplify the Boltzmann treatment by making an expansion in large
$q^0$ in \Eq{eq:Boltzmann_symbols}:
\begin{equation}
\left[ -iq^0 + \hat{\cal C} \right]^{-1}
 = \frac{i}{q^0} + \frac{1}{(q^0)^2} \hat{\cal C} + \ldots \,.
\end{equation}
In other words, we can treat $iq^0 \chi$ as the dominant contribution
in \Eq{eq:Boltzmann_full} and solve the equation iteratively.
The leading term, $\chi=iT/q^0$, is
pure imaginary and does not contribute to the spectral weight.  The next
contribution arises by substituting the leading contribution into the
collision term, and gives
\begin{eqnarray}
Q_a \chi^{(2)}_a = \frac{T}{(q^0)^2 f_0(1{-}f_0) 4p\nu_a} 
\hspace{-0.15in}&& \int_{\p'\k\k'} \sum_{bcd}
	|{\cal M}^2| (2\pi)^4\delta^4(P{+}K{-}P'{-}K')
        f(p) f(k)
\\ && \times
        [1{\pm}f(p')][1{\pm}f(k')]
\left[ Q_a+\hat\p\cdot \hat\k Q_b - \hat\p\cdot \hat\p' Q_c
  -\hat\p\cdot \hat\k' Q_d \right] \,, \nonumber
\end{eqnarray}
which, substituted into \Eq{eq:rho_j}, yields a spectral weight of
\begin{eqnarray}
\frac{\rho}{q^0} & = & \frac{2}{(q^0)^2 T}
\int_{\p\p'\k\k'} \frac{1}{2}\sum_{abcd} |{\cal M}^{ab}_{cd}|^2
	(2\pi)^4\delta^4(P{+}K{-}P'{-}K')
	f(p)f(k)(1{\pm}f(p')(1{\pm}f(k'))
\nonumber \\ && \hspace{1in} \times
	Q_a \hat\p \cdot \left(
	Q_a \hat\p + Q_b \hat\k - Q_c \hat\p' - Q_d \hat\k' \right)
\nonumber \\
 & = & \frac{1}{(q^0)^2 T}
\int_{\p\p'\k\k'} \frac{1}{4}\sum_{abcd} |{\cal M}^{ab}_{cd}|^2
	(2\pi)^4\delta^4(P{+}K{-}P'{-}K')
	f(p)f(k)(1{\pm}f(p')(1{\pm}f(k'))
\nonumber \\ && \hspace{1in} \times
	\left(Q_a \hat\p + Q_b \hat\k - Q_c \hat\p' - Q_d \hat\k' \right)^2
	\,,	
\label{rho_smallomega}
\end{eqnarray}
where the last expression is a symmetrization of the previous one over
the indices $abcd$.

This expression has a simple diagrammatic interpretation.  The momentum
integration, particle sum, matrix element, momentum conserving delta
function, and population function give the total rate per 4-volume for a
scattering process to occur.  The factor of $1/4$ corrects for
overcounting $a\leftrightarrow b$ and $c\leftrightarrow d$ (or is a
symmetry factor if these labels are the same).
A factor of $e^2$ times the $1/(q^0)^2$ times the
final factor represent the probability that the scattering process
should lead to an off-shell photon emission with frequency $q^0$.

\begin{figure}
\centerbox{0.6}{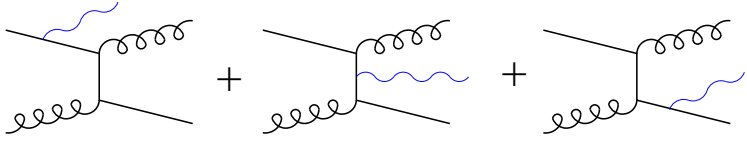}
\caption{\label{fig_Compton}
The amplitude for a Compton scattering to produce a photon involves a
sum over where the external photon attaches.}
\end{figure}

To see this, consider how a typical scattering rate is modified by the
addition of a soft photon.  For concreteness consider Compton
scattering.  The photon can attach to any charged external leg or
internal propagator, see Figure \ref{fig_Compton}.  However, it is only
the amplitude when it attaches to an external leg which is $1/q^0$
enhanced, as we now discuss.

Consider the diagram where a soft photon is emitted from one of
the quark external legs.  For this discussion we write the photon
momentum as $R$ to avoid confusion with the charge.  Take the external
state momentum to be $P$, so the momentum
on the internal propagator is $L=P\mp R$ (where $\mp$ is $-$ for initial
and $+$ for final state particles.)
Because the particle is on shell, $P^2=0$. 

Writing the polarization vector for the photon as $\epsilon$,
the diagram with the photon emission on an initial state leg 
differs from the diagram without any emission by the substitution
\begin{equation}
u(P) \rightarrow  \frac{-\nott{L}}{L^2}\,[-Q \nott{\epsilon}]\, u(P)
\label{eq:sub}
\end{equation}
and in a shift to the kinematics of the rest of the process,
$P\rightarrow L$.  Because $R$ is small, the change to the kinematics of
the rest of the diagram represent a small correction and can be
neglected. 

Since $L=P-R$, 
\st
\frac{\not{L}}{L^2} = \frac{{\nott P}-{\nott R}}{P^2 + R^2 -2P\cdot
    R} \approx \frac{\nott P}{-2P\cdot R} = \frac{\nott P}{2p^0 r^0}
 \,.
\stp
Anticommuting $\nott{P}$ with $\nott{\epsilon}$ and using that 
$\nott{P} u(P)=0$ (the Dirac equation), \Eq{eq:sub} is
approximately
\begin{equation}
u(P) \rightarrow  \frac{Q}{r^0}\; \frac{2P\cdot \epsilon}{2p^0} \; u(P)
= \frac{1}{r^0} \; Q \hat\p \cdot \epsilon \; u(P)\, .
\end{equation}
The same thing happens when we consider an outgoing external leg, except
that it is $+2P\cdot R$ rather than $-2P\cdot R$ which appears, so the
overall sign is opposite.  The term where the photon attaches to an
internal leg is proportional to the virtuality of that
leg, which is at least $O(gT)$ and is therefore subdominant at small
frequency.  Therefore the amplitude for the \brem\ process is identical
to the amplitude for the regular
$\twotwo$ diagram, but with an overall multiplicative factor of
$\frac{-eQ}{r^0}
  \sum_{\rm in-out} \hat \p \cdot \epsilon$.
On squaring and summing over polarization states (and there are three,
since the photon is off-shell and in the pure time direction), we
recover the factor $e^2/(q^0)^2 \times (\sum_{\rm in-out}Q\hat\p)^2$
which we found above.

Therefore the result, \Eq{rho_smallomega}, is just what we would obtain
diagrammatically by considering bremsstrahlung processes, after
accounting for the fact that the photon kinematics are soft.

With the exception of Coulomb scattering, all $2\leftrightarrow 2$
processes in the plasma have a rate $\sim \alphas^2 T^4$ up to logs.
Coulomb scattering is faster by a factor of $1/\alphas$, but only
because of small angle processes for which $(\hat\p - \hat\p')^2 \sim
\alphas$; therefore its contribution to \Eq{rho_smallomega} is only
logarithmically larger than a typical $2\leftrightarrow 2$ process.
Namely, each such process gives a contribution of order
\begin{equation}
\frac{\rho}{q^0} \sim \frac{\alphas^2 T^3}{(q^0)^2} \, ,
\qquad
\frac{dW}{d^4 q} \sim \frac{\alpha^2 \alphas^2 T^3}
	{(q^0)^4} \, ,
\end{equation}
which is parametrically $O(\alpha^2/T)$ for $q^0 \sim gT$.  

Our result is
the same order, and shows the same $q^0$ dependence, as the term
found by BPY.  However, the diagrams they
include in their analysis only give rise to
photon production from Compton scattering and its crossings,
not, for instance, from Coulomb scattering.  Therefore their calculation
must be incomplete.

Let us make a more detailed comparison.  The complete result of BPY, for
general quark number and charge, is \cite{BPY}

\begin{eqnarray}\label{DileptonHTL}
\frac{dW(q)}{d^4q}\bigg|_{\p=0} & = & \frac{2\alpha^2\sum Q^2}
{\pi^4 (q^0)^2}
\int_0^{\infty} k^2 dk\int^\infty_{-\infty}d\omega
\int^\infty_{-\infty} d\omega' n_f(\omega) n_f(\omega')
\delta(q^0-\omega - \omega ') 
\nonumber\\ && \hspace{1in}
\times \Bigg\{4\left(1- \frac{\omega^2-\omega'^2}{2k q^0}\right)^2
\rho_+(\omega,k) \rho_-(\omega',k) 
\nonumber\\ && \hspace{1.1in}
+\left(1+ \frac{\omega^2+\omega'^2-2k^2-2m^2}{2k q^0}\right)^2
\rho_+(\omega,k) \rho_+(\omega',k)
\nonumber\\ && \hspace{1.1in}
+\left(1- \frac{\omega^2+\omega'^2-2k^2-2m^2}{2k q^0}\right)^2
\rho_-(\omega,k) \rho_-(\omega',k)
\nonumber\\ && \hspace{1.1in}
+\Theta(k^2-\omega^2)
\frac{m_q^2}{4k(q^0)^2}\left(1-\frac{\omega^2}{k^2}\right)
\nonumber\\ && \hspace{1.2in}
\times \left[\left(1+\frac{\omega}{k}\right)\rho_+(\omega',k)
  +\left(1-\frac{\omega}{k}\right)\rho_-(\omega',k)\right]\Bigg\}
\, ,
\end{eqnarray}
where
\st
\rho_\pm(\omega,k) =\frac{\omega^2 - k^2}{2m_q^2}
[\delta(\omega-\omega_\pm(k)) + \delta(\omega+\omega_\mp(k))] +
\beta_\pm(\omega,k)\Theta(k^2-\omega^2) \, ,
\stp
and
\st
\beta_{\pm}(\omega,k) = \frac{(m_q^2/2k)(1\mp\omega/k)} {\{ \omega \mp k -
  (m_q^2/k)[Q_0(\omega/k) \mp Q_1(\omega/k)]\}^2 + [(\pi
    m^2_q/2k)(1\mp\omega/k)]^2} \, .
\stp
We are interested in the small $q^0$ behavior, for which 
$\omega' \simeq -\omega$.  In this case the prefactors on the spectral
weights contain large $1/(q^0)^2$ enhancements, but only when 
$\omega^2 \neq k^2$, meaning that only
the $\beta_{\pm}$, or cut, piece of $\rho_{\pm}$ contributes
appreciably.  In this case the integrand behaves as $m_q^4/k^4 q_0^2$, 
indicating a large $k$ log divergence, which BPY remarkably did not
comment on.  This is the familiar log divergence due to the $s/t$ nature
of the Compton cross-section, seen from the infrared side.

Attempting only to compute the coefficient of this log divergence, we
may expand in $|\omega|,k \gg m_q$.  (Here $m_q^2=[\cf=4/3]g^2 T^2/8$ is
the effective thermal mass of a quark.)  The expansion is relatively
straightforward and leads to
\begin{equation}
\frac{dW}{d^4 q}(\q=0) = \frac{2\alpha^2 \sum Q^2}{\pi^3 q_0^2}
\; \frac{3 m_q^4}{16\pi q_0^2} \int \frac{dk}{k} \int_{-k}^k
\frac{d\omega}{k} \left[ 1-\frac{\omega^2}{k^2} \right] \,.
\label{eq:BPYexpand}
\end{equation}
The corresponding log divergent coefficient in the spectral weight is
\begin{equation}
\frac{\rho(\q=0,q^0)}{q^0} = 12 \sum Q^2 \times \frac{m_q^4}{4\pi q_0^2 T}
\ln \frac{T}{m_q} \, ,
\label{eq:BPYresult}
\end{equation}
where we have taken the upper and lower limits of the $k$ integration
to be $T$ (where population functions enter) and $m_q$ (where the full
expression is required and the result is smaller).  It is not clear to
us how this divergence was handled in the numerical work presented by
BPY \cite{BPY}.

Now we approach the same calculation using \Eq{rho_smallomega}.  We will
show as claimed that the BPY result coincides with the contribution from
Compton type processes.  The
relevant summed matrix element is $12\sum Q^2 \times 8 \cf^2 s/t$, with
$s,t$ the usual Mandelstam variables.  Here $12$ is the number of
degrees of freedom per quark flavor (2 spins times 3 colors times
particle/antiparticle) and $8\cf^2 s/t$ is the matrix element squared
summed on all other species' spins and summed over Compton and
annihilation processes.  The phase space integrations can be reduced
using the technique of \cite{AMY6}, to give
\begin{eqnarray}
\frac{\rho}{q^0}& = & \frac{12 \sum Q^2 \times 8 \cf^2}{2^8 \pi^5 q_0^2 T}
\int_0 dk \int_{-k}^k d\omega \int_0^{2\pi} \frac{d\phi}{2\pi} \int_0 dp_1
\int_0 dp_2 \; \frac{s}{t}
\nonumber \\ && \hspace{1.2in} \times
f_f(p_1) [1{+}f_b(p_1)]
f_f(p_2) [1{+}f_b(p_2)]
\times (\hat\p_1 - \hat\p_2)^2 \, ,
\end{eqnarray}
where we have already made the approximation $\omega,k \ll T$ in writing
the limits of $p$ integrations, population functions, and the vectorial
term.  Physically, $k,\omega$ represent the exchange momentum and energy
in a Compton type process and $p_1,p_2$ represent the energies of the
two participating (anti)quarks.  The Mandelstam variable ratio and the
required angle are given
approximately by 
\st
\frac{s}{t} \simeq \frac{2p_1 p_2}{k^2} (1{-}\cos\phi) \, , \qquad
\hat\p_1 \cdot \hat\p_2 = 1-\frac{s}{2p_1p_2}
\simeq 1-\frac{k^2{-}\omega^2}{k^2}(1{-}\cos\phi) \,.
\stp
The $p_1$, $p_2$, and $\phi$ integrations are then completely
straightforward, and they yield
\st
\frac{\rho}{q^0} = 12\sum Q^2 \frac{3 m_q^4}{16\pi q_0^2 T}
\int \frac{dk}{k} \int_{-k}^k \frac{d\omega}{k} 
  \left[ 1-\frac{\omega^2}{k^2} \right] \, ,
\stp
which is identical to \Eq{eq:BPYexpand} and leads to the same result for the
spectral weight as \Eq{eq:BPYresult}.  

\begin{figure}
\centerbox{0.85}{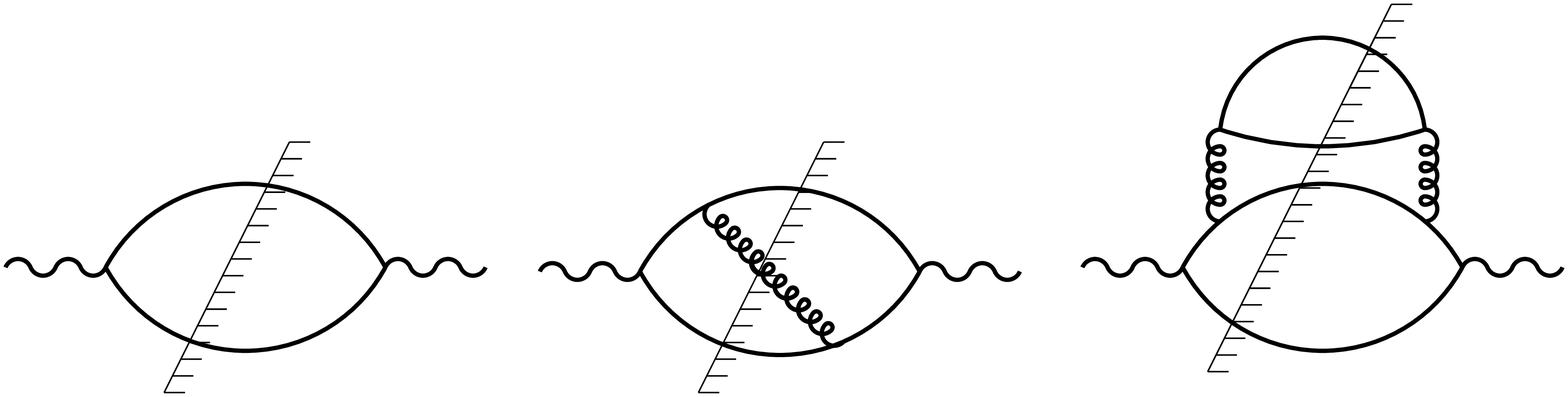}
\caption{\label{fig2} 
One loop diagram considered by BPY, a two-loop
diagram, and an example three-loop diagram which we find contributes
at the same order for $q^0 \sim gT$.  The hashed diagonal line
represents a cut; lines crossing the cut are put on-shell.}
\end{figure}

We see that the leading small
$q^0$ behavior in the BPY calculation corresponds to evaluating just the
Compton-type processes in the kinetic theory approach.  Their
calculation is therefore obviously incomplete.
What went wrong in the BPY calculation?  Their mistake was to assume
that the loopwise expansion would be valid for photon production,
modulo HTL corrections.  This would be true were they considering a
hard ($q^0 \sim T$) virtual external photon--in this case the first
corrections are known and are $O(\alphas)$ \cite{Majumder}.  However,
considering the $q^0 \sim gT$ regime, low-loop calculations can be
suppressed by kinematic restrictions which disappear at higher loop
orders.  For instance, consider the 1-loop diagram analyzed by BPY, a
two-loop diagram, and an example of the 3-loop diagrams which we find to
dominate at small momentum, all shown in Fig.\ \ref{fig2}.
Besides the explicit $e^2$ in the one-loop diagram, the fermion lines
are kinematically constrained to carry momentum $\sim q^0 \sim gT$ so
that they may both be simultaneously on-shell; since the photon line
attaches to this low-momentum line, the diagram has an extra factor of
$q_0^2 \sim \alphas$ and is $O(\alpha \alphas)$.  This is not
enhanced by any Bose stimulation functions because the soft lines are
fermionic.  

For the three-loop diagram, on the other hand, there is an explicit
factor of $e^2 g^4$, but the softness of the incoming momentum makes no
restrictions on the sizes of the three loop momenta, which can each be
$O(T)$ and at large angles with respect to each other without preventing
the four cut lines from being simultaneously on-shell.  The momentum
factors therefore lead to a $T^2$ behavior.
Furthermore, there are two uncut, almost on-shell
propagators which lead to a $1/(P\cdot Q)^2 \sim 1/g^2$ enhancement,
giving a rate which is
$O(\alpha \alphas)$, the same order as the 1-loop diagram.  
This is true even if {\em all} of the lines in the 3-loop diagram are
hard--the need to add loops is not just the well-understood need to
resum HTL's on soft lines.  

Almost the
same is true of the 2-loop diagram, except that if the lines are all
hard then there is a collinear restriction for the three cut lines to be
simultaneously on-shell, reducing the phase space by $O(\alphas)$ and
again making the rate $O(\alpha \alphas)$.  Higher loop diagrams are
suppressed with respect to these, however, because loop factors continue
to appear but no additional kinematic constraints are lifted, since
there are no remaining kinematic constraints on the 3-loop diagrams.
This argument fails when $q^0 \sim g^4 T$, at which point a
large class of diagrams begin to contribute (which we effectively
resummed in using kinetic theory).

\section{Kinetic Theory evaluation of spectral weight}

We are now ready to proceed with the evaluation of
\Eq{eq:Boltzmann_full} and \Eq{eq:rho_j}.
This involves solving a non-trivial integral
equation, which cannot be handled analytically.  We must revert to a
numerical treatment, which can however be made highly accurate by using
the variational techniques applied to conductivity in \cite{AMY1,AMY6}.

The quantities of interest in the evaluation of
\Eq{eq:Boltzmann_first}, $Tf_0(1{-}f_0) \hat{\p}\equiv S(\p)$ and
$\hat\p \chi(p)$, are general vector functions of $\p$ (with particularly
simple angular dependence).  The collision integral and $q^0
f_0(1{-}f_0)$ can be considered as linear operators on such functions
(with $q^0 f_0(1{-}f_0)$ being an almost trivial operator).  It is
useful to consider such functions of $\p$ as elements in the vector
space of all functions of $\p$ with the same angular dependence, and to
define an inner product on such functions $f$ and $g$, namely
\st
\Big(f,g\Big) \equiv \frac{1}{T^3}\sum_a \nu_a \int \frac{d^3\p}
{(2\pi)^3}f_i(\p)g_i(\p)\, .
\stp
The utility of this inner product is that the collision operator is
real, positive, and symmetric under this inner product, and
the quantity we want, \Eq{eq:rho_j}, is the inner product of two
functions, $\Big(\chi,S \Big)$.  Therefore, a variational approach for
determining $\chi(\p)$ will have errors which are quadratic in the
quality of the variational \ansatz.

With this in mind, we define the functional
\st
Q[\chi]\equiv\Big(\chi,S \Big)
-\frac{1}{2}\Big(\chi,\left[\mathcal{C}-iq^0 f_0(1{-}f_0)
\right]\chi \Big)\,.
\stp
Extremizing $Q[\chi]$ with respect to $\chi$ precisely yields the
linearized Boltzmann equation (\ref{eq:Boltzmann_symbols}), and the
extremal value of $Q$ is simply related to the quantity we want;
\st
\frac{\rho}{q^0} = \frac{2}{T} \Re \Big( \chi, S \Big)
= 4 \Re Q_{\rm{extr}} \,.
\stp
Unfortunately, the matrix $\mathcal{C}-iq^0 f_0(1{-}f_0)$ is not
positive definite, so the real part of the extremal value is not
necessarily a lower bound for the result at finite $q^0$.

The extremization of the functional $Q[\chi]$ may be performed in
several ways.  We will follow \cite{AMY1,AMY6} and treat the problem
variationally. For an exact solution, we would need to work with an
infinite dimensional space of arbitrary functions $\chi(p)$.  However,
by choosing a suitable set of basis functions, ${\phi^{(m)}(p)}$, and
performing the maximization in this finite dimensional variational
vector space, we are able to get a highly accurate approximate
result. We will consider the \ansatz,
\st
\chi^a(p) = \sum^K_{m=1}\tilde\chi^a_m\phi^{(m)}(p)\, ,
\stp
where $K$ is the size of the basis set considered and $\tilde\chi^a_m$
are the variational parameters used to tune the maximization of
$Q[\chi]$.  Inserting our \ansatz\ into the source and collision terms
of the functional yields linear and quadratic combinations of the
arbitrary coefficients $\tilde\chi^a_m$,
\begin{eqnarray}
\Big(\chi,S\Big) & = & \sum_{a,m}\tilde\chi^a_m\tilde S^a_m \, ,
\\
\Big(\chi,\cal{C}\chi\Big) & = &
\sum_{a,m}\sum_{b,n}\tilde\chi^a_m \tilde C^{ab}_{mn} \tilde\chi^b_n
\, ,
\\
\Big( \chi,q^0 f_0(1{-}f_0)\chi \Big) & = & 
\sum_{a,m} \sum_{b,n} \tilde\chi^a_m \tilde q^0_{mn}\delta_{ab}
\tilde\chi^b_n \,.
\end{eqnarray}
In these expressions, $\tilde S^a_m \equiv \Big( S^a,\phi^{(m)}\Big)$,
$\tilde q^0_{mn} \equiv q^0 \Big( \phi^{(m)},
f_0(1{-}f_0)\phi^{(n)}\Big)$, and
$\tilde C^{ab}_{mn} \equiv \Big(\phi^{(m)} ,
\mathcal{C}^{ab} \phi^{(n)}\Big)$. 
Restricted to this subspace, and treating $\tilde\chi$, $\tilde S$ as
vectors and $\tilde {\cal C}$ and $\tilde q^0$ as matrices with index
$(a,m)$, the functional becomes
\st
\tilde Q[\tilde\chi]= \tilde\chi^\trans \tilde S
-\half\tilde\chi^\trans [\tilde C -i\tilde q^0] \tilde\chi \, .
\stp

We note that the previous equation has both a real and imaginary part
to it.  Since we ultimately want the real part,
it will be convenient to write the complex structure in matrix form;
\begin{equation}
	Q[\tilde \chi] \equiv
	\Biggl(
	    \Big[ \tilde S \,, 0 \Big] ,
	    \left[ \begin{array}{c} \Re\,\tilde\chi \\ \Im\,\tilde\chi
	\end{array} \right]
	\Biggr)
	-
	\frac 12
	\Biggl(
	    \Big[ \Re\, \tilde\chi \,, \Im \, \tilde\chi \Big] ,
	    \left[
		\begin{array}{cc} \tilde{\cal C} & \tilde q^0 \\
		\tilde q^0 & -\tilde{\cal C} \\ \end{array}
	    \right]
	    \left[ \begin{array}{c} \Re\,\tilde\chi \\ 
		\Im\,\tilde\chi \end{array} \right] 
	\Biggr)
	\, .
\end{equation}

Finding the approximate form of the spectral weight (the real part of
the extremum) is now an elementary linear algebra exercise and is given by
\begin {equation}
    Q_{\rm max} = \half\,{\cal S}^{\rm T} \, {\cal M}^{-1} \, {\cal S} \,,
\label{eq_Q_symbols}
\end {equation}
where,
\begin {equation}
    {\cal S} \equiv
    \left[ \begin{array}{c} \tilde{S}_n\vphantom{\Big(}\\
			0 \end{array} \right] , \qquad
    {\cal M} \equiv
	\left[ \begin{array}{cc}
	    \tilde{C}_{mn} & \omega_{mn} \vphantom{\Big|}\\ 
	    \omega_{nm}\vphantom{\Big|} & -\tilde{C}_{mn}
	\end{array} \right] \, .
\end {equation}

The next step is to choose suitable trial functions.
As discussed in \cite{AMY6}, a good choice for the basis
functions  will consists of
\st \label{choice}
\phi^{(m)}(p) = \frac{(p/T)^{m-1}}{(1+p/T)^{K-2}}, \qquad m =
  1,...,K.
\stp
This set spans both the function $S$ itself, and the variational
\ansatz, $\chi(p)\propto p$, which has been found to work remarkably
well for conductivity \cite{Heiselberg}.
These functions are not orthogonal, but they do not
need to be.  At any finite $K$ they are not complete, but they span
nested subspaces (the space spanned at one $K$ sits in the space spanned
at a larger $K$), and as $K \to \infty$, the basis in fact becomes
complete.  The integrations to establish $\tilde{S}_n$ can be performed
as straightforward quadratures integrations; and while the integrals for
$\tilde{C}_{mn}$ are less straightforward, they can be reduced to
4-dimensional integrals and also carried out by quadratures.  One
complication is that even after scaling out the explicit factor of $g^4$
from the collision term $\tilde C$, it retains a logarithmic dependence
on $g$ because Compton and Coulombic scattering processes give infrared
logarithmic divergences, cut off by plasma screening; and the plasma
screening cutoff scale varies with $g$.  Both our approach to the
numerical integrations, and the details of how screening cut off IR
singular scattering processes are dealt with 
at length in \cite{AMY6}, so we will not present the details
again here; the main new feature is the complex matrix structure which
we have already clarified.

\section{Results}

\begin{figure}
\centerbox{0.7}{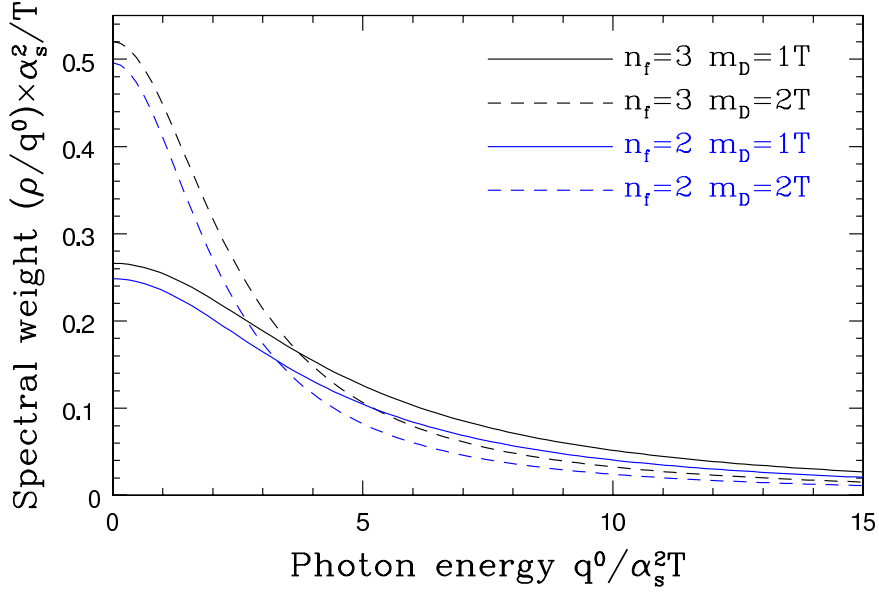}
\caption{\label{fig_result1}
Spectral weight vs.\ $q^0$ with the dominant $\alphas$ dependence
factored out.  For $\nf=2$, the two values of $\mD$ correspond
to $\alphas=0.06$ and $0.24$; for $\nf=3$ they correspond to
$\alphas=0.05$ and $0.21$.}
\end{figure}

The scale where the spectral weight shows nontrivial behavior is
$q^0 \sim g^4 T$, since this is where the collision term, which is
explicitly $O(g^4)$, becomes comparable to $q^0$ in evaluating
\Eq{eq_Q_symbols}.  In this case $\rho/q^0 \sim T/g^4$.  The
weak-coupling behavior of the spectral weight, scaling out this explicit
behavior, is displayed in Fig.\ \ref{fig_result1}.  The result depends
on the number and charge of quark flavors, and it also retains some
dependence on the coupling, because of the screening
masses mentioned above.  The simplest functional form with the same
behavior as $\rho/q^0$--an even function with a finite value at zero and
$1/q_0^2$ tails--is the Lorentzian.  Though we have not shown it, the
large $\mD$ (large coupling) curves in Fig.\ \ref{fig_result1} are quite
well fitted by a Lorentzian, but the small $\mD$ curves are less so.

\begin{figure}
\centerbox{0.7}{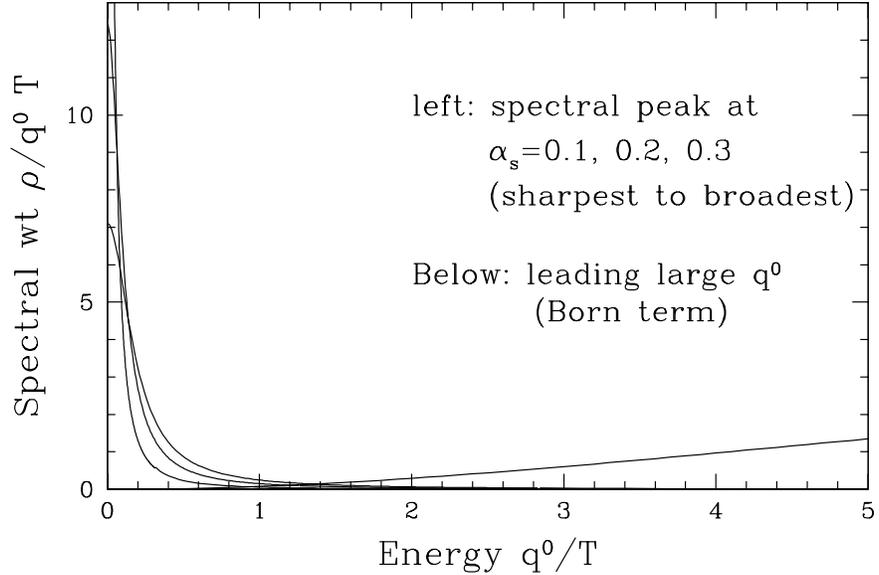}
\caption{\label{fig_result2}
Dominant large $q^0$ and small $q^0$ contributions to the spectral
weight.  The rising curve is the leading-order Born term; the three
curves sharply peaked at the origin are our results for a 3-flavor
plasma with $\alphas=0.1,0.2,0.3$.  As
the coupling is reduced, the peak at small $q^0$ becomes taller and
narrower, but the area underneath is unchanged.  The Born result is
valid where it is much larger than the peak, and the peak is valid where
much larger than the Born term.  In the overlap region
(where $q^0\sim gT$) the detailed behavior is not known.}
\end{figure}

We also present the functional form of $\rho/q^0$ as a function of $q^0/T$
for 3-flavor QCD in Fig.\ \ref{fig_result2}.  The figure shows the small
$q^0$ behavior we have found for three values of the coupling $\alphas =
0.1,0.2,0.3$, together with the leading-order large $q^0$ behavior.  The
region where the curves cross is parametrically $gT$; neither
calculation works here.  The calculation of BPY was intended to
establish $\rho/q^0$ here, but we have already shown that it is
incomplete, so at present there are no leading-order perturbative
results in this region.

\begin{figure}
\centerbox{0.7}{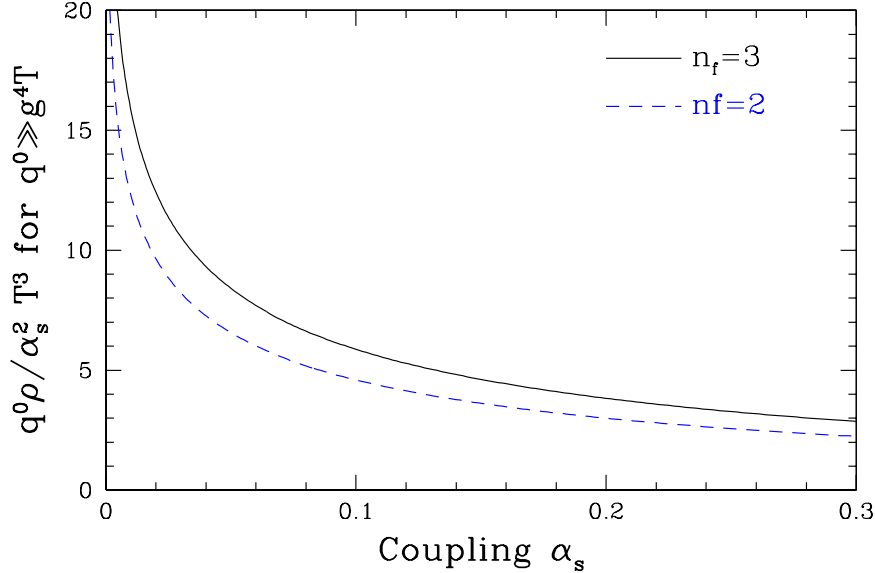}
\caption{\label{fig_result3}
Large $q^0$ asymptotic of the kinetic treatment (which is the small
$q^0$ asymptotic of the $q^0\sim gT$ treatment) as a function of the
gauge coupling, for $\nf=2,3$.}
\end{figure}

As we already emphasized, it is especially easy to extract the $q^0 \gg
g^4 T$ asymptotic of the above curves, which is supposed to agree with
the $q^0 \ll gT$ asymptotic of BPY.  We obtained an explicit expression
in \Eq{rho_smallomega}. We present this asymptotic as a
function of $\alphas$ in Fig.\ \ref{fig_result3}.  It can also be
expanded in the logarithm of the coupling; since it is the collision
operator and not its inverse which appears, this expansion terminates
rather than containing all orders in inverse logs, as is the case for
the conductivity \cite{AMY6}.  Explicitly,
\begin{eqnarray}
\label{eq:NLL}
\frac{\rho}{q^0} \left[ g^4 T \ll q^0 \ll gT \right]
& = & \frac{\alphas^2 T^3 \sum Q^2}{q_0^2} 
\left[ \frac{4\pi}{3} + \frac{16 (6+\nf)}{3\pi} \right]
\left( \ln\frac{1}{g} + C \right)
\, , \\
C & = & \left\{ \begin{array}{cc} -0.010 & \nf=2 \\
-0.039 & \nf=3 \\ \end{array} \right. \, . \nonumber
\end{eqnarray}
The first term in the square bracket arises from Compton-type processes,
the second is 
from Coulombic scattering; they were obtained by evaluating
\Eq{rho_smallomega} using the techniques of section IV of \cite{AMY1}.
We saw in \Eq{eq:BPYresult} that the treatment of BPY reproduces the
first but not the second term.  Unfortunately, the range of validity of
this expansion is very narrow, since it becomes negative before
$\alphas=0.1$; the leading-log treatment does particularly poorly here.
Physically this is because the dominant momentum of a charge carrier is
small, and at physically relevant couplings the typical scattering
process changes its direction substantially.  The case is worse than for
transport phenomena because they are dominated by larger momentum
particles.

Fig.\ \ref{fig_result2} shows that the spectral weight has a peak near
$q^0=0$.  As the coupling increases, the peak spreads out but becomes
lower.  But even at quite large coupling, $\alphas = 0.3$, the peak is
still very close to zero frequency.  Further, the area under the peak
remains the same regardless of the value of $\alphas$, because the
small $q^0$ result we have obtained obeys a {\sl sum rule}.
To see this, consider the integral over all $q^0$ of $\rho/q^0$;
\st
\int dq^0 \; \frac{\rho}{q^0} = \int dq^0 \;\Re 
2 \Big( S, [{\cal C} - iq^0 f_0(1{-}f_0)]^{-1} S \Big) \, .
\stp
Inserting a complete set of eigenfunctions $\Lambda$ of the operator
{\cal C}, with  eigenvalue $\lambda$, this integral becomes
\begin{eqnarray}
\int dq^0 \; \frac{\rho}{q^0} & = & 
\sum_\Lambda \int dq^0 \; 2 \Re \Big( S \,,\, 
  [{\cal C}-iq^0 f_0(1{-}f_0)]^{-1} \Lambda \Big)\;
  \Big( \Lambda\,,\, S\Big)
\nonumber \\
& = &
\sum_\Lambda \int dq^0 \; 2 \Re \Big( S \,,\, 
  [\lambda-iq^0 f_0(1{-}f_0)]^{-1} \Lambda \Big)\;
  \Big( \Lambda\,,\, S\Big) \,.
\end{eqnarray}
The operator ${\cal C}$ has been replaced by a number $\lambda$.
Therefore, inside the integral implicit in the inner product, the
inverse, $[\lambda-iq^0 f_0(1{-}f_0)]^{-1}$, is an ordinary (rather than
operator) inverse, and there is no obstacle to performing the $q^0$
integration.  The integral is straightforward and yields
\st
\int dq^0 \; \frac{\rho}{q^0} =
\sum_\Lambda \Big( S \, , \, \frac{2\pi}{f_0(1{-}f_0)} 
\, \Lambda \Big) \Big( \Lambda\, ,\, S\Big)
= 2\pi \Big( S\,,\, \frac{1}{f_0(1{-}f_0)} S\Big) \,.
\stp
Explicitly, the remaining integral is
\st
2\pi \Big( S\, ,\, \frac{1}{f_0(1{-}f_0)} \, S\Big)
= 2\pi \sum_a \nu_a Q_a^2 \frac{1}{T^3}\int \frac{d^3p}{(2\pi)^3}
f_0(p) \left(f_0(p) - 1 \right) = \frac{2\pi \nc \sum_q Q_q^2}{3} \, .
\stp

\section{Discussion}

We have presented a leading-order calculation of the electromagnetic
current-current spectral weight, and therefore of the dilepton
production rate, in weakly coupled, thermal QCD and for dilepton
energies $q^0 \ll gT$.  In this regime, the computation of the spectral
weight requires applying kinetic theory.  There is a simple
interpretation of the kinetic theory result for $q^0 \gg g^4 T$ in terms
of bremsstrahlung from scattering processes in the plasma.

Our results show that the asymptotic behavior
found by Braaten, Pisarski, and Yuan, $\rho/q^0 \propto q_0^{-2}$ for
$q^0 \ll gT$, is correct down to the scale $q^0 \sim g^4 T$, though the
coefficient which they found for this behavior is in error.  This error
was caused by an insufficiently careful power-counting; a wider class of
diagrams than they considered contribute at leading order, because the
one-loop diagram has kinematic constraints which disappear at higher
loop order.

At the scale $q^0 \sim g^4 T$, the spectral weight's behavior moderates,
so that $\lim_{q^0\rightarrow 0} \rho/q^0$ is finite, as is required by
the finiteness of the electrical conductivity.  The weakly coupled
spectral weight shows a large peak at small $q^0$, but the area under
this peak is finite.

Furthermore, there is a sum rule satisfied by $\rho/q^0$; its $q^0$
integral over the aforementioned peak is independent both of the
coupling and of the details of the scattering processes which determine
its detailed form.
The significance of this sum rule is that it makes it very hard to use
Euclidean correlation functions to reconstruct the
peak value, $\lim_{q^0\rightarrow 0} \rho(q^0)/q^0 = 6 \sigma/e^2$
(with $\sigma$ the electrical conductivity).  Since this is precisely
what must be done to find the conductivity from lattice gauge theory
calculations, we believe that (at least at weak coupling) such a
reconstruction is extremely challenging.

To clarify this last point, we refer to the discussion of
Aarts and Resco \cite{AartsResco}.  The Euclidean correlator at zero
spatial momentum and finite Euclidean time $\tau$ is determined in terms
of the spectral weight by an integral,
\st
G_{_{\rm E}}(\tau) = \int \frac{dq^0}{2\pi} \frac{\rho(q^0)}{q^0}
K(\tau,q^0) \, ,
\qquad
K(\tau,q^0) = \frac{q^0 \cosh[q^0(\tau-\beta/2)]}
              {\sinh(\beta q^0/2)} \, .
\stp
The point is that $K(\tau,q^0)$ is essentially flat and $\tau$
independent near
$q^0=0$, which means that $G_{_{\rm E}}$ at any $\tau$ essentially
captures the area under the small $q^0$ peak.  But this area is fixed by
the sum rule and carries no information about the detailed shape of the
peak.  Hopefully, at very large coupling the peak becomes broad enough
(or even disappears \cite{Teaney_N4SYM}) that the weak $q^0$ dependence
in $K(\tau,q^0)$ is enough to see its structure--though even at $\alphas
= 0.3$, the edge of the range where we believe perturbative methods, the
peak remains very narrow, with most of its support at $q^0 < 0.5 T$.
So long as the peak is narrow, fitting will be
difficult, especially without an \ansatz\ for the shape of the peak.
Fortunately, as we mentioned, the peaks shown in Fig.\ \ref{fig_result1}
for large coupling are surprisingly well fit by a Lorentzian, exactly
the fitting form advocated by Aarts and Resco \cite{AartsResco}.

\medskip

\centerline{\bf Acknowledgements}

\medskip

The authors are indebted to Gert Aarts and Derek Teaney for useful
conversations, and to Peter Petreczky, for conversations and for
spurring us to carry out this study.  We apologize to Braaten, Pisarski
and Yuan for sounding too negative about their work (which in fact we
think was an excellent paper).  This work was partially supported by
grants from the National Science and Engineering Research Council of
Canada (NSERC) and by 
le Fonds Qu{\'e}b{\'e}cois de la Recherche sur la Nature et les
Technologies (FQRNT).


\end{document}